\documentclass[reviewcopy]{elsart}

\usepackage{amsmath}
\usepackage{amsfonts}
\usepackage{amssymb}
\usepackage [ansinew] {inputenc}
\usepackage[dvips]{graphicx}
\usepackage[nomarkers,nolists]{endfloat}
\usepackage[square, comma, sort&compress]{natbib}

\usepackage{theorem}
\theorembodyfont{\upshape}

\newcommand{\N} {
\ensuremath{\mathcal{N}}
}

\newcommand{\J} {
\ensuremath{\mathcal{J}}
}

\begin{document}

\begin{frontmatter}
\title{Correct thermodynamic forces in Tsallis Thermodynamics: connection with
Hill Nanothermodynamics}
\author{Vladimir~Garc\'{\i}a-Morales\corauthref{cor1}}
\ead{vladimir.garcia@uv.es}, \author{Javier~Cervera},
\author{Julio~Pellicer} \corauth[cor1]{Corresponding author. Tel: +34
96 354 3119; Fax: +34 96 354 3385}
\address{Departament de Termodin\`{a}mica, Universitat de
Val\`{e}ncia, C/Dr. Moliner 50, E-46100 Burjassot, Spain}
\begin{abstract}
\noindent{The equivalence between Tsallis Thermodynamics and
Hill's Nanothermodynamics is established. The correct
thermodynamic forces in Tsallis thermodynamics are established.
Through this connection we also find a general expression for the
entropic index $q$ which we illustrate with two physical examples,
allowing in both cases to relate $q$ to the underlying dynamics of
the Hamiltonian systems.}
\end{abstract}
\end{frontmatter}

\section{Introduction}

During the last decade there has been growing interest in Tsallis
(nonextensive) thermostatistics \cite{Tsallis88, web}. This
formalism has been successfully applied to a wide variety of
statistical systems at scales ranging from particle and nuclear
physics \cite{Navarra03} to astrophysics \cite{Torres97} and
situations involving long range interactions \cite{Long} as well
as low dimensional maps and multifractals \cite{multifractals}.
Another important success is the straightforward explanation of
the occurrence of Levy distributions in nature \cite{Levy}. The
building block of Tsallis thermostatistics, known as Tsallis
entropy, was introduced in $1988$ \cite{Tsallis88}
\begin{equation}
\mathcal{S}^{*}=\frac{\sum_{j}^{\Omega}p_{j}^{q}-1}{1-q} \qquad (q
\in \Re) \label{entropy}
\end{equation}
Here $p_{j}$ is the probability of the microstate $j$ and
$\mathcal{S}^{*}$ is a generalized entropic measure which reduces
to the Gibbs-Shannon one $S=-k\sum^{\Omega}_{j}p_{j}\ln p_{j}$
when the entropic parameter $q$ tends to unity. The main feature
of this entropic form is its nonadditivity. Given two subsystems
$A$ and $B$ of a composite nonextensive system $A+B$ it can be
easily checked \cite{Tsallis88} that
\begin{equation}
\mathcal{S}^{*}(A+B)=\mathcal{S}^{*}(A)+\mathcal{S}^{*}(B)+
(1-q)\mathcal{S}^{*}(A)\mathcal{S}^{*}(B) \label{nonadditivity}
\end{equation}
Despite the wide variety of applications that the formalism has
proven \cite{web} there are some important points concerning
Hamiltonian systems that have not yet been elucidated. Among
these, the problem of how can $q$ be calculated for these systems
has attracted great attention \cite{qmeaning} and is, arguably,
the main opened issue of Tsallis thermodynamics (TT) \cite{Cho02}.
Another point that has not been clarified is what is the correct
form of the thermodynamic forces in TT \cite{Vives02, AbeToral,
critics}. It has also been stated that Tsallis entropy cannot have
a well defined thermodynamics because of its nonadditivity
\cite{Salinas97, critics}.

Quite interestingly also, there exist already another previous
important generalization of traditional thermostatistics which is
well grounded physically and which was originally conceived for
dealing with nonextensive systems \cite{Chamberlin02}. This
formalism is Hill's Nanothermodynamics \cite{founds} (NT) and was
originated in 1962 generalizing the concept of Gibbs' chemical
potential to complex nonextensive systems that can vary its
entropy by fragmentating in smaller subsystems. NT has been
applied succesfully since then to ferromagnetism
\cite{Chamberlin00}, glassy systems \cite{Chamberlin99} and
liquid-vapor interfaces \cite{Hill98, GarciaMorales03}. NT is well
grounded physically from its very beginning in contrast with TT
which relies in the \emph{ad hoc} (although beautiful) postulate
for the entropy inspired in the multifractal formalism
\cite{Tsallis88}. TT has, however, many practical advantages and
allows to deal quite straigthforwardly with complex systems
(specially those exhibiting power laws statistically since these
can be closely described by means of Tsallis q-exponential
distributions). NT could be misunderstood as applicable only to
small sizes or small numbers of particles. But stars, clusters and
multifractals can be considered also as NT systems (the range of
the interactions and/or correlations are of the order of the size
of the system or longer) and, in general, all systems in which
their equilibrium properties depart from the standard description,
which relies in the concept of extensivity \cite{Chamberlin02}. NT
is a rigorous statistical foundation of thermodynamic finite-size
effects as well as others coming from the complexity of the
interactions. It is therefore interesting questioning whether if
there is any connection between TT and NT.

In this article we clarify all these points. Specifically: i) We
establish the equivalence between TT and NT, ii) this connection
allows us then to introduce an adequate form for the thermodynamic
forces in TT, overcoming previous difficulties \cite{critics} iii)
the nonadditivity property is then understood for complex systems
thermodynamically and, importantly, iv) we provide an expression
for $q$ that can be used to \emph{calculate} this quantity in
Hamiltonian many particle systems relating it to its internal
dynamics.

\section{Formal equations of Tsallis Thermodynamics (TT)}

We present first previous rigorous results of TT and NT to better
clarify our approach. Maximization of the Tsallis entropy imposing
the constraints $\langle X_{\alpha}  \rangle_q = \sum_{j}^\Omega
p_{j}^{q} X_{\alpha}(j)/\sum_{j}^\Omega p_{j}^{q}$ for the biased
average $\langle X_{\alpha}  \rangle_q$ of an extensive quantity
$X_{\alpha}$ leads to recover a Gibbs-like differential equation
for the entropy
\begin{equation}
d  \mathcal{S}^{*}  =  \sum_{\alpha}  y_{\alpha}   d  \langle
X_{\alpha} \rangle_q, \label{termo}
\end{equation}
The rest of the formal equations of TT have been recently
established \cite{Vives02}. The Euler equation of TT is
\begin{equation}
\sum_{\alpha} \langle X_{\alpha}  \rangle_q y_{\alpha} =
\frac{1}{1-q} \left [ 1+ (1-q)  \mathcal{S}^{*} \right ] \ln \left
[  1+ (1-q) \mathcal{S}^{*} \right ]. \label{Euler}
\end{equation}
and the Gibbs-Duhem equation has the form
\begin{equation}
\sum_{\alpha} \langle  X_{\alpha} \rangle_q dy_{\alpha} =  \ln
\left [ 1+ (1-q) \mathcal{S}^{*} \right ] d\mathcal{S}^{*}
\label{GibbsDuhem}
\end{equation}
A reasonable condition (local additivity) that variables $\langle
X_{\alpha} \rangle_q$ must satisfy and which is used in the
derivation of Eqs. (\ref{Euler}) and (\ref{GibbsDuhem}) is $
\left.d \langle  X_{\alpha}(\lambda A)
\rangle_q/d\lambda\right|_{\lambda=1}= \langle X_{\alpha}(
A)\rangle_q \label{condition} $ for a system $\lambda$ times
bigger than another $A$. Note that this by no means imply that
$\langle  X_{\alpha}(\lambda A) \rangle_q=\lambda\langle
X_{\alpha}(A) \rangle_q$ (global additivity) \cite{Vives02}, since
$\lambda$ is replaced by unity. Global additivity generally holds
only when an ensemble of $\lambda$ independent noninteracting
systems is considered (see below) but not necessarily upon
subdivision of a system into smaller fractions (it is necessary,
for example, to provide some energy to separate two droplets $A$
and $B$ from a bigger droplet $A+B$ because of the energetic cost
of creating additional interfaces). Tsallis entropy satisfies
\cite{Vives02} $\left.d \mathcal{S}^{*}(\lambda A)
/d\lambda\right|_{\lambda=1}=\left [ 1+ (1-q) \mathcal{S}^{*}(A)
\right ] \ln \left [  1+ (1-q) \mathcal{S}^{*}(A) \right ]/(1-q).$
It is to be noted that, as occurs with their respective
counterparts in standard thermodynamics, Eqs. (\ref{termo}) to
(\ref{GibbsDuhem}) are not independent, i.e. knowing two of them
the other one can be trivially deduced. The above general
equations and the Legendre transform mechanism allows to determine
all thermodynamic quantities of a nonextensive system.

\section{Hill Nanothermodynamics (NT)}

Let us now introduce Hill's formalism of NT. It is based in
maximization of the Gibbs-Shannon entropy but contains a new
entropic thermodynamic potential, the subdivision entropic
potential (SEP) $\J$, which couples nonlinearly the natural
thermodynamic variables of the system. Introduction of this new
potential is made at the ensemble level, in which a set of
$\mathcal{N}$ identical, noninteracting small systems is
considered \cite{founds}. In this case, the set of formal
equations are $\J
d\N+\sum_{\alpha}y_{\alpha,H}d\left<X_{\alpha}\right>_{H,t}=dS_{t}$
(Gibbs); $\J
\N+\sum_{\alpha}y_{\alpha,H}\left<X_{\alpha}\right>_{H,t}=S_{t}$
(Euler) and $\N
d\J+\sum_{\alpha}\left<X_{\alpha}\right>_{H,t}dy_{\alpha,H}=0$
(Gibbs-Duhem). Here, subindex $H$ means ``Hill's variables'' which
are the \emph{physical} (averaged) locally extensive
($\left<X_{\alpha}\right>_{H,t}$) and intensive ($y_{\alpha,H}$)
ones. $t$ means "total" and denotes the properties of the whole
ensemble. Thus $S_{t}$ is the total entropy for the ensemble of
$\mathcal{N}$ systems. We have $S_{t}=\N S$ and
$\left<X_{\alpha}\right>_{H,t}=\N \left<X_{\alpha}\right>_{H}$ in
terms of the respective variables for one system (note that here
the members of the ensemble do not interact directly although, of
course, can exchange heat, volume and particles depending on the
ensemble considered, this being itself a formal construction in
which total properties are $\N$ times those of one system). The
formal equations of NT become then for one system \cite{founds}
\begin{eqnarray}
\sum_{\alpha}y_{\alpha,H}d\left<X_{\alpha}\right>_{H}&=&dS \label{HGi}\\
\sum_{\alpha}y_{\alpha,H}\left<X_{\alpha}\right>_{H}&=&S-\J \label{HEuler}\\
\sum_{\alpha}\left<X_{\alpha}\right>_{H}dy_{\alpha,H}&=&-d\J
\label{HGD}
\end{eqnarray}

\section{Connection between TT and NT}

Eqs. (\ref{HGi}) to (\ref{HGD}) are to be compared with Eqs.
(\ref{termo}) to (\ref{GibbsDuhem}) respectively. It can be seen
that if we make the following connection $S \equiv
\mathcal{S}^{*}$, $y_{\alpha,H} \equiv y_{\alpha}$,
$\left<X_{\alpha}\right>_{H} \equiv \left<X_{\alpha}\right>_{q}$
and
\begin{equation}
\J \equiv \mathcal{S}^{*}-\left[1+(1-q)\mathcal{S}^{*}\right]\frac{\ln \left[1+(1-q)\mathcal{S}^{*}\right] }{1-q}
\label{def}
\end{equation}
the structure of the formal equations of TT and NT is the same.
Here the Tsallis entropy $\mathcal{S}^{*}$ is not only the
physical one: its property of nonadditivity, see Eq.
(\ref{nonadditivity}), is also the basis for the SEP $\J$ which is
found to be necessary to explain the thermal behavior of small
systems (at least). TT describes, thus, the most general thermal
equilibrium, the nanothermodynamic equilibrium
\cite{Chamberlin03}, in which the new potential $\J$ plays a
decisive role. $\J$ vanishes for a macroscopic system, for which
one has also $q=1$ in Eq.(\ref{def}), and is a measure of the
(thermodynamic) smallness of the system. In making this
correspondence it could seem counterintuitive equating an additive
entropy $S$ to a nonadditive one $\mathcal{S}^{*}$. However, $S$
is Gibbs-Shannon like because Hill's entropy is written
statistically in terms of thermal averages considering a
nanosystem in equilibrium with a macroscopic heat bath at a given
average temperature. The nonlinear coupling of the intensive
variables is accounted for thermodynamically in Eq.(\ref{HEuler})
by means of the generalized potential $\J$. This is totally
equivalent to consider the nanosystem at the same average inverse
temperature arising this time not from coupling to a macroscopic
heat bath but to a neighbouring environment with fluctuating
temperature following a chi-square distribution (without need of
introducing $\J$ in this case). The latter leads to Tsallis
statistics \cite{qmeaning, Wilk}, the entropy being nonextensive
and the probability distributions depending on the parameter $q$
(that comes from the order of the chi-square (Gamma) distribution
considered). Some authors have shown that the Tsallis distribution
is the Laplace transform of the chi-square distribution
\cite{qmeaning, Wilk}. This is the key point in passing from
Boltzmann-Gibbs distributions to Tsallis q-exponential
distributions.

Nonadditivity in the entropy (a statistical approach) and
introduction of the potential $\J$ (a thermodynamic approach) are
only different pathways that lead to the same thermodynamics. As
shown below, thermodynamic excess functions (Hill's NT) are linked
to entropic nonadditivity (Tsallis TT). It is also to be noted
that the ensemble approach followed by Hill to derive the
thermodynamics of a small system is equivalent to assume that
local additivity of parameters and Eq. (\ref{HGi}) hold as well as
the following expression for the additivity breaking of the
entropy $\left.d S(\lambda
A)/d\lambda\right|_{\lambda=1}=S(A)-\J.$

Despite the different statistical behavior both Hill and Tsallis
entropies can be made to coincide numerically and the differences
of both formalisms concern only the way of averaging. It is worthy
insisting on the fact that thermodynamically speaking, however,
these differences are irrelevant, since the structure of the
formal equations is the same. Both formalisms, TT and NT, preserve
the Gibbsian structure of thermodynamics \cite{Tsallis88, founds}.
Fluctuations are also gaussian-like in both formalisms
\cite{founds, Vives02}. Furthermore, if the range $0<q<1$ is
considered to be the physically meaningful one \cite{Vives02}
fluctuations in TT are larger for $q \ne 1$ than in the extensive
case \cite{Vives02} which is again consistent with the fact that
fluctuations in NT for a system with $\J \ne 0$ are also larger
than in the $\J=0$ case \cite{founds}. Although Tsallis
probability distributions are more general than Boltzmann-Gibbs
ones, it is important to note that, in NT, $\J$ introduces
additional degrees of freedom in the evaluation of the relevant
partition function \cite{Chamberlin00} (which in NT is always
related to excess quantities besides the traditional thermodynamic
ones). Furthermore, and quite importantly, both TT and NT share
the property of ensemble non-equivalence for $q \ne 1$ and $\J \ne
0$, respectively.

\section{Correct form for the Lagrange parameters}

It has been questioned the physical meaning of the parameters
$y_{\alpha}$, and some authors \cite{Vives02, AbeToral} have
concluded previously that the physical quantities of interest are
${\hat y}_{\alpha} = y_{\alpha}/[1+(1-q) \mathcal{S}^{*}].$ This
conclusion, however, is wrong. Parameters ${\hat y}_{\alpha}$ are
naturally related to a thermodynamic formalism in which use of
Renyi entropy (and \emph{not} Tsallis entropy) is made. As a
function of Tsallis entropy, the Renyi entropy
$\mathcal{\widehat{S}}^{*}$ has the form
$\mathcal{\widehat{S}}^{*} =  \ln \left [1+ (1-q) \mathcal{S}^{*}
\right]/(1-q)$ \cite{Tsallis88}. However, while Tsallis entropy is
nonextensive and stable \cite{AbeW}, Renyi entropy is extensive
but thermodynamically unstable \cite{Lesche82} and cannot be
employed to generalize standard thermodynamics \cite{AbeW}.
Parameters ${\hat y}_{\alpha}$ also arise when it is assumed
\emph{ad hoc} global additivity of energy and the maximum Tsallis
entropy for two systems brought into contact is considered
\cite{AbeToral}. This, however, is also incorrect, since energy
cannot be globally additive where entropy does not satisfy this
property (otherwise, temperature could not be an intensive
variable \cite{critics}). It is to be noted therefore that neither
the correct thermodynamic forces nor the physical meaning of the
formal relations were previously established. Through the above
correspondence and from Hill's NT, it is now known that the
$y_{\alpha}$ must be equal for two different systems put in
contact at equilibrium. \emph{The $y_{\alpha}$ are then the
physically meaningful} thermodynamic forces. This allows to
establish equilibrium TT, which is now free from recent criticisms
\cite{critics} arising from the use of incorrect thermodynamic
forces.

\section{Excess quantities. Meaning of the nonadditivity of
Tsallis entropy}

Excess thermodynamic quantities $\langle
X_{\alpha}\rangle_q^{(x)}$, as considered, for example, in surface
physics \cite{founds, Rowlinson82}, can be defined as $\langle
X_{\alpha}(A,B) \rangle_q^{(x)}= \langle X_{\alpha}(A+B)
\rangle_q-\langle X_{\alpha}(A) \rangle_q-\langle X_{\alpha}(B)
\rangle_q$ By applying Eq. (\ref{termo}) to the composite system
$A+B$ and then to each system $A$ and $B$ separately and using
these definitions, we obtain $ d
[(1-q)\mathcal{S}^{*}(A)\mathcal{S}^{*}(B)] = \sum_{\alpha}
y_{\alpha} d \langle X_{\alpha} \rangle_q^{(x)}$. This clearly
leads to quantify the entropy excess in TT as
$\mathcal{S}^{*(x)}=(1-q)\mathcal{S}^{*}(A)\mathcal{S}^{*}(B)$
\cite{GarciaMorales04b}. We see, thus, that TT provides also a
specific microscopic statistical model for the thermodynamic
excess entropy and this is the very physical meaning of the
nonadditivity property. Each independent system $A$ or $B$ is
analogous to a bulk phase. Despite their statistical independency,
a thermodynamic coupling exists however between both systems. This
can be modelled as an interface separating them and contributing
with an excess entropy $\mathcal{S}^{*(x)}$ to the properties of
the total system $A+B$. A nonextensive system in TT can hence be
understood as a two-phase-like system at the vicinity of the
critical point (since the nonadditivity property holds at all
scales which means having always significant interfaces separating
clusters of $A$ from clusters of $B$). It is interesting that
Eq.(\ref{def}) can be recasted as $\J =
\mathcal{S}^{*}-\mathcal{\widehat{S}}^{*}-(1-q)\mathcal{S}^{*}\mathcal{\widehat{S}}^{*}$
which is also similar to the Tsallis entropy scheme of
nonadditivity, see Eq.(\ref{nonadditivity}). The SEP $\J$
represents the net balance between fragmentation or aggregation of
the system \cite{founds, Chamberlin00, Chamberlin99}. This SEP is
hence to be viewed in TT as a thermodynamic force which can be
understood \emph{formally} in terms of an equilibrium of the
actual physical nonextensive system (with physical entropy
$\mathcal{S}^{*}$) with its completely fragmentated and
uncorrelated state in which the system is reduced to a gas of its
noninteracting constituents. The gas has entropy
$-\mathcal{\widehat{S}}^{*}$. The minus sign comes from the fact
that this gas represents disintegration, against the cohesion (the
positive Tsallis physical contribution) which preserves the
integrity of the nonextensive system. This equilibrium is made
through an interface contributing with
$-(1-q)\mathcal{S}^{*}\mathcal{\widehat{S}}^{*}$. Note that Renyi
entropy $\mathcal{\widehat{S}}^{*}$ is extensive, and hence it
describes well this completely fragmentated extensive dilute gas
(which is an unstable and supersaturated one) of the physical
system.

It can be seen also that $\J$ and $\mathcal{S}^{*}$ are related
differentially as $d\J = -(1-q)\mathcal{\widehat{S}}^{*}d
\mathcal{S}^{*}$. This equation is somewhat unusual and has no
counterpart in standard thermodynamics. If we define $f \equiv d\J
/d\mathcal{S}^{*}$ we see that $f(\lambda A)=\lambda f(A)$. This
means that the chance to aggregate depends linearly on the size of
the system, which is quite reasonable.

\section{General expression for q and examples}

By using our definition of excess entropy, we obtain
\begin{equation}
1-q=\frac{\mathcal{S}^{*(x)}}{\mathcal{S}^{*}(A)\mathcal{S}^{*}(B)}=
\frac{S^{(x)}}{S(A)S(B)} \label{defq}
\end{equation}
since $\mathcal{S}^{*}(A)\equiv S(A)$ and $\mathcal{S}^{*}(B)
\equiv S(B)$ through our connection while Eq. (\ref{HGi}) holds
also in NT for the composite system $A+B$ as well as for each
separate system $A$ and $B$. In Eq. (\ref{defq}) all the
quantities appearing in the r.h.s can be always \emph{calculated}
from NT, which means that there is a way to know $q$ from first
principles estimations. We now illustrate this assertion by
calculating $q$ for nonpolar, nonstructured liquids (as argon) at
the vicinity of the critical point. Liquid clusters develop inside
the supersaturated vapor phase at all scales. From critical point
theory \cite{Rowlinson82} it is well known that the surface
tension $\sigma$ scales at the vicinity of the critical point
($T_{r}\to 0$) with the reduced temperature $T_{r} \equiv
(1-T/T_{c})$, where $T_{c}$ is the critical temperature, as $\sim
\sigma_{0}T_{r}^{\mu}$ (with $\mu \approx 1.26$ in three
dimensions, and $\sigma_{0}$ being the surface tension amplitude).
The correlation length of phase $i$ ($i=A, B$), $\xi_{i}$, scales
as $\sim \xi_{0i}T_{r}^{-\mu/(D-1)}$ \cite{Rowlinson82} where $D$
is the dimension and $\xi_{0i}$ the amplitude of the correlation
length. In this limit, the bulk entropies can be written in terms
of the critical bulk entropic density $s_{c}$ as $S_{i}=s_{c}V_{i}
\approx s_{c}\xi^{D}_{i}$, where $V_{i}$ are the volumes occupied
by each phase. The excess entropy $S^{(x)}$ is equal to
$-ad\sigma/dT$ \cite{founds} where $a \approx \xi^{D-1}$ is the
surface area. Replacing all these quantities (taking $A$ as the
liquid phase) in Eq. (\ref{defq}) and considering $D=3$ we obtain
\begin{equation}
1-q=\frac{\mu}{4\pi\omega_{c}s_{c}^{2}\xi_{0A}^{3}\xi_{0B}^{3}}
T_{r}^{3\mu-1} \label{defq2}
\end{equation}
where we have introduced the critical wetting parameter
$\omega_{c}=kT_{c}/(4\pi\xi_{0A}^{2}\sigma_{0}) \approx 0.78$
which is a universal constant for all fluids
\cite{Fisher92,GarciaMorales03}. We have shown recently that
$\omega_{c}$ \emph{is related directly to the dynamics of the
critical clusters} \cite{GarciaMorales03}. Concretely, a value
$\omega_{c}=0.78$ means that the critical clusters behave
\emph{universally} as a combination of translation and vortex
rotational motion \cite{GarciaMorales03}. This calculation then
relates the entropic parameter $q$ to the dynamical quantity
$\omega_{c}$ and the universal exponent $\mu$ that can be obtained
separately from renormalization group theory \cite{Rowlinson82}.
It is to be noted that although renormalization group theory is an
exact approach given the Hamiltonian, real systems exhibit
nonideal features due to the complexity of the interactions that
can lead to departures from the translationally invariant
hamiltonians usually considered in these cases. This is
particularly dramatic when studying ferromagnetism
\cite{Chamberlin00, Chamberlin03}. In our specific example,
classical scaling laws are used only here as an approximation. It
is to be noted that far from the critical point (always following
the coexistence line) $q$ does not necessarily departs from unity
since the bulk entropies do not scale with the correlation
lengths. Eq. (\ref{defq2}) holds only approximately as the
critical point is approached (as assumed from the scaling laws
above used). It is to be noted that at the critical point $q=1$.
This is consistent with the fact that the two phases in
coexistence merge then into a single macroscopic homogeneous one
provided that the size of the clusters tend to infinity. This
means that, in addressing clusters, the thermodynamic limit is
taken at the critical point ($N \to \infty$ and $V \to \infty$)
and hence standard thermodynamics should be regained as is the
case here. The r.h.s. of this equation is vanishingly small for
non-structured simple fluids, as argon, because of the large
numerical values for $s_{c}$. For these fluids, then, $q \approx
1$ and this justifies the success of the Gibbs method of the
dividing surface within standard thermostatistics in regarding
these systems. Departures from this ideal behavior are observed,
however, for complex structured liquids as water
\cite{Rowlinson82} in which there exists anomalies due to the
presence of hydrogen bonding and dipole orientation which
drammatically complicates the statistical description. The
presence of structuredness reduce the bulk entropy of the liquid
and, furthermore, interfacial effects strongly increase (as it is
known from the value of the surface tension of water which
surprisingly high when compared to simple fluids
\cite{Rowlinson82}). It is then to be expected $q$ to depart
significantly from unity in liquid water.

There are many other systems as Ag nanoparticles for which surface
functions can be significantly higher than bulk values
\cite{Nanda03} which would lead, from Eq.(\ref{defq}), to values
of $q$ significantly different from unity. Ferromagnetic materials
have been also proven to significantly depart from standard
thermostatistics \cite{Chamberlin00}, and this has been
interpreted succesfully in terms of a nanothermodynamic
equilibrium \cite{Chamberlin00, Chamberlin03}.

Another example of a formal two-phase-like system is a correlated
ionic liquid in the vicinity of a strongly charged macroion
\cite{GarciaMorales04}. In this situation the behavior of the
correlated counterions at the vicinity of the macroion is quite
different to those far from the macroion (constituting a
Poisson-Boltzmann uncorrelated and disordered liquid)
\cite{Netz01, GarciaMorales04}. For weak coupling, the correlated
liquid is still a 3D fluid-like one but the rescaled interfacial
area available to each molecule is affected by correlations and is
approximately given by $2\Xi$ (where $\Xi$ is the coupling
parameter entering in the Hamiltonian) \cite{Netz01}. It is clear,
then, that in the weak coupling regime ($0<\Xi<1$, $q \lesssim 1$)
and from Eq. (\ref{defq}), $1-q \approx c\Xi$, where $c$ is a
small constant (since $S^{(x)}$ depends linearly in the
interfacial area). This is consistent with the previously fitted
curve for $q(\Xi)$ since in the weak coupling limit we have
$1-q=1-1/(1+0.091\Xi)^{0.68} \approx 0.062\Xi$
\cite{GarciaMorales04}.

\section{Concluding remarks}

We have established a connection between TT and NT which provides
a sound and unambiguous physical basis for TT. As a bonus, it has
allowed us to justify that the $y_{\alpha}$'s are the correct
physical thermodynamic forces (difficulties pointed out in
\cite{critics} are now overcomed) and has led us to two
expressions that can be used quantitatively to evaluate the degree
of nonextensivity in Hamiltonian systems. These are Eq.
(\ref{def}), which introduces the connection between both
formalisms relating thermodynamic smallness to nonextensivity, and
Eq. (\ref{defq2}), which allows for an straightforward evaluation
of $q$ from NT. A nonextensive system with $q \ne 1$ can now be
understood in the thermodynamic limit as one composed of two
entangled phases in a critical regime in which the (fractal)
interface separating them, besides other finite size effects,
contributes significantly to the total entropy at all scales.

\section*{Acknowledgements}

We acknowledge financial support from the MCYT (Ministry of
Science and Technology of Spain) and the European Funds for
Regional Development (FEDER) under project No. MAT2002-00646.

\end{document}